\documentstyle[11pt,epsfig]{article}

\def\ra{\rightarrow}

\def\be{\begin{eqnarray}}
\def\ee{\end{eqnarray}}
\def\bq{\begin{equation}}
\def\eq{\end{equation}}
\def\ben{\begin{enumerate}}\def\een{\end{enumerate}}

\def\mn{m_N}
\def\kF{k_F}

\def\F1{F_1}
\def\ft1{{\tilde{F}_1}}
\def\ft1omega{\tilde{F}_1^\omega}
\def\tr{{\rm tr}}
\def\prl {Phys. Rev. Lett.}\def\pr{Phys. Rev.}
\def\np{Nucl. Phys.}\def\pl{Phys. Lett.}
\def\la{\langle}\def\ra{\rangle}

\def\A0{A_0}
\def\Jvec{\bm{J}}

\def\pb{\mbox{\boldmath$ p$}}\def\qb{\mbox{\boldmath $q$}}
\def\Bp{\mbox{\boldmath$ p$}}\def\Bq{\mbox{\boldmath $q$}}
\def\Bk{\mbox{\boldmath$ k$}}\def\Br{\mbox{\boldmath $r$}}
\def\Bs{\mbox{\boldmath $s$}}
\def\qbhat{\mbox{\boldmath $\hat{q}$}}\def\taub{\mbox{\boldmath $\tau$}}

\def\kbhat{\mbox{\boldmath $\hat{k}$}}
\def\nubhat{\mbox{\boldmath $\hat{\nu}$}}

\def\sigmab{\mbox{\boldmath $\sigma$}}
\def\Bnabla{\mbox{\boldmath $\nabla$}}
\def\Bsigma{\mbox{\boldmath $\sigma$}}

\def\Beta{\mbox{\boldmath $\eta$}}
\def\roughly#1{\mathrel{\raise.3ex\hbox{$#1$\kern-.75em%
\lower1ex\hbox{$\sim$}}}}
\def\bm#1{\mbox{\boldmath$#1$\/}}

\def\lsim{\roughly<}
\def\gsim{\roughly>}

\renewcommand{\thefootnote}{\fnsymbol{footnote}}
\begin{document}
\setlength{\baselineskip}{20.6 pt}
\begin{titlepage}
\hfill{SNUTP 98-061}
\vspace{.3cm}
\begin{center}
\ \\
{\Large \bf Scaling Of Chiral Lagrangians And Landau Fermi Liquid Theory
For Dense Hadronic Matter}

\ \\
\vspace{0.2cm}
 {Bengt Friman$^{(a)}$, Mannque Rho$^{(b,c)}$ and Chaejun Song$^{(d)}$ }

{\it (a) Theory Group, GSI, Planckstr. 1, D-64291 Darmstadt, Germany \&\\
\vskip -0.2cm Institut f\"ur Kernphysik, TU Darmstadt, D-64289 Darmstadt,
Germany}\\ \vskip -0.2cm
    {\small (E-mail: b.friman@gsi.de)}

{\it    (b) Service de Physique Th\'eorique, CEA Saclay,\\ \vskip -0.2cm
     F-91191 Gif-sur-Yvette, France}\\ \vskip -0.2cm
{\small (E-mail: rho@spht.saclay.cea.fr)}

{\it    (c) School of Physics, Korea Institute for Advanced Study,\\ 
\vskip -0.2cm Seoul 130-012, Korea}

{\it    (d) Department of Physics and Center for Theoretical Physics\\
\vskip -0.2cm Seoul National University, Seoul 151-742, Korea}\\ \vskip -0.2cm
{\small (E-mail: chaejun@fire.snu.ac.kr)}
\vskip 0.3cm
\end{center}
\vskip 0.5cm

\centerline{\bf ABSTRACT}
\vskip 0.6cm
\noindent
We discuss the Fermi-liquid properties of hadronic matter derived
from a chiral Lagrangian field theory in which Brown-Rho (BR) scaling is
incorporated. We identify the BR scaling as a contribution to
Landau's Fermi liquid fixed-point quasiparticle parameter from
``heavy" isoscalar meson degrees of freedom that are integrated out
from a low-energy effective Lagrangian. We show that for the vector
(convection) current, the result obtained in the chiral Lagrangian
approach agrees precisely with that obtained in the
semi-phenomenological Landau-Migdal approach. This precise agreement
allows one to determine the Landau parameter that enters in the effective
nucleon mass in terms of the constant that characterizes BR scaling.
When applied to the weak axial current, however, these two approaches
differ in a subtle way. While the difference
is small numerically, the chiral Lagrangian approach implements
current algebra and low-energy theorems associated with the axial
response that the Landau method misses
and hence is expected to be more predictive.

\end{titlepage}
\setcounter{footnote}{0}
\renewcommand{\thefootnote}{\arabic{footnote}}

\topmargin -0.7in
\section{Introduction}
\indent\indent
At very low energies, the relevant degrees of freedom for strong
interactions in nuclear matter are pions, nucleons and other
low-mass hadrons identified in the laboratory and the appropriate
theory is an effective quantum field theory involving these hadrons
even though the fundamental theory is known to be QCD with quarks
and gluons\cite{weinberg}. Just which and how many hadronic degrees
of freedom must appear in the theory depends upon the energy scale
that is probed. Thus for example, if one is probing energies of a
few MeV's as in the case of low-energy properties of two-nucleon
systems, then the nucleon field as a matter field and possibly
pions as pseudo-Goldstone bosons would suffice.

In this paper, we would like to extend the strategy of effective
field theories to many-body systems and a density regime
corresponding to a shorter-length or higher-energy scale than that
probed by the low-energy two-nucleon systems \cite{PKMR,pkmr-EFT}.
This would entail two important changes to the effective
Lagrangian: First we need to introduce
 more massive degrees of freedom (such as vector mesons
and/or higher-dimensional operators in the nucleon fields)
in the effective Lagrangian and second, we
need to take into account the Fermi sea of nucleons in the bound system.

The principal aim in this paper is then to tie in together various results
obtained previously in diverse contexts into a unified framework so
as to be able to extrapolate our ideas into the kinematic domains
that are yet to be explored experimentally. In doing this,  we
shall be using nonrelativistic arguments which are justified for
low-energy and low-density processes we are concerned with here. A
relativistic formulation more appropriate for high-energy and
high-density heavy-ion processes is in progress and will be
presented elsewhere.

The basic strategy we will develop is as follows. First we will
present an argument for an effective chiral Lagrangian which in the
mean field approximation corresponds to a non-topological soliton
describing a lump of nuclear matter. The parameters of this
effective Lagrangian will then be identified with the fixed-point
quantities in Landau Fermi liquid theory. Given this
identification, one can associate certain mean field quantities of
heavy mesons (e.g., the light-quark vector mesons $\rho$ and
$\omega$) to BR scaling via Landau parameters. We first illustrate
how this chain of arguments works for electromagnetic properties of
heavy nuclei. Turning the arguments around, we determine the BR-scaling 
parameter $\Phi$ at nuclear matter density from magnetic
moments of heavy nuclei in terms of the Landau parameter
$F_1^\omega$ associated with massive isoscalar vector meson degrees
of freedom that are integrated out from the effective Lagrangian.
We then use a similar line of arguments to derive the corresponding
formulas for the axial current. In this paper, we shall focus on
the processes that are dominated by pionic effects, that is, those
to which the ``chiral filter mechanism" \cite{KDR} applies, namely,
the electromagnetic convection current and the axial charge
operator.

This paper is organized as follows. In Section \ref{eft}, effective field
theories that figure in nuclear physics are described including a brief
summary of Landau's Fermi liquid theory adapted to strongly interacting
nuclear systems. The calculation of the electromagnetic current for a particle
sitting on top of the Fermi sea in Landau-Migdal theory
and in chiral Lagrangian theory is given in Section \ref{em}. The Landau
parameter figuring in the nucleon effective mass is determined in terms of
the parameter of the chiral Lagrangian that scales as a function of density
(a la BR scaling).
The problem of treating axial charge transitions in heavy nuclei is presented in
Section \ref{axial}. The two methods -- Fermi liquid and chiral Lagrangian --
are found to give almost same numerical results at nuclear matter density
but differ in a subtle way due to the intricacy with which chiral
symmetry is manifested in nuclear systems. This is discussed in Section
\ref{comparison} where the electromagnetic current and the weak current are
compared. BR scaling that enters in the chiral Lagrangian is tested and checked
phenomenologically in Sections \ref{kaon} and
\ref{test}. These two sections provide support for the
basic assumption that goes into the link between Fermi-liquid theory and
chiral Lagrangian theory. What we have established is summarized in Section
\ref{discussions} wherein some unresolved/open problems are mentioned.
In Appendix A, we show how to compute relativistically
the pionic contribution to the Landau
parameter $F_1$ using a Fierz transformation. Appendix B sketches how the
particle-hole graph figuring in the ``back-flow" argument is computed.
Remarks not directly relevant to to the theme of the paper but helpful for
the discussions are relegated to footnotes.

\section{Effective Field Theories (EFT)}\label{eft}
\indent\indent
There are two superbly {\it effective} effective field theories for nuclear
physics. One is chiral Lagrangian field theory for low-energy nonperturbative
description of hadrons and the other is Landau Fermi liquid theory applied to
nuclear matter.
\subsection{EFT for dilute systems}
\indent\indent
For two-nucleon systems at very low energies considered in
\cite{PKMR,pkmr-EFT}, one can integrate out all meson degrees of
freedom including the pions and set the cutoff near one-pion mass.
One then writes an effective Lagrangian in terms of the nucleon
field in a systematic (chiral) expansion to compute the irreducible
graphs, while summing the infinite set of reducible diagrams to
describe the deuteron bound state and the scattering state with a
large scattering length. Since the system is dilute, the parameters
of the theory can be taken from free-space (zero-density)
experiments. In principle, we should be able to calculate these
parameters  from QCD but at present we do not know how to perform
this calculation in practice. The results in \cite{PKMR,pkmr-EFT}
confirm that the approach works remarkably well. When the pion
field is included in addition, it provides a ``new" degree of
freedom and improves the theory even further and  allows one to go
higher in energy scale~\cite{pkmr-EFT}. This can be formulated
systematically in terms of chiral perturbation theory.

But what about heavier (denser) nuclei or higher energy scales?

\subsection{EFT for dense systems}
\indent\indent
In going to heavier many-nucleon systems, the standard approach has
been to start with a Lagrangian whose parameters are defined in
free space and then develop perturbative and non-perturbative
schemes to account for the complex dynamics involved. Higher energy
scales will be involved since the interactions between nucleons in
such systems sample all length scales and hence other
degrees of freedom than nucleonic and pionic need be introduced. In doing
such calculations, symmetry constraints, such as those of chiral
symmetry, are found to be useful but not always properly implementable.
Basically phenomenological in character, given a sufficient number
of free parameters, such an approach can be quite successful but
one cannot check unambiguously
that it is consistent with the modern notion of
effective field theory. As such, it is difficult to gauge the power
of the theory. When something does not work, then there is very
little one can do to improve on it since there is no systematic
strategy available.

In this paper we will take a different route. Following Lynn
\cite{lynn}, we shall assume that a high-order (in chiral counting)
effective chiral action supports a non-topological soliton solution
that corresponds to a chiral liquid with a given baryon number A.
Lynn proposes to construct such an effective action using chiral
perturbation theory {\it to all orders of the chiral expansion}, but
up to now explicitly deriving such an action has not been feasible.
Lacking such a first-principle derivation, we propose to develop an
effective Lagrangian strategy applied to dense many-body systems
resorting to certain assumptions based on symmetries which are to
be justified a posteriori. Given such an action possessing a stable
non-topological soliton, we follow Lynn's proposal to identify
such a soliton solution as the ground state of a heavy nucleus and
to make fluctuations around that ground state. Excitations on top
of that state could then be described in terms of {\it the
parameters determined at that minimum}, the bulk properties of
which are to be generically characterized by the density of the
state that is probed. There have recently been several works along this
line. For instance, Furnstahl et al.\ \cite{furnstahl} construct such
an effective action consisting of ``heavy baryons'' (nucleons) and
heavy mesons using arguments based on the ``naturalness condition''
of chiral symmetry of QCD and show that in the mean field the
effective Lagrangian quantitatively describes the ground state of
nuclear matter as well as the excitation spectra of finite nuclei.
The point pertinent to us in the work of Furnstahl et al.\ is
that their formulation is basically equivalent to a variant of
Walecka mean-field theory. A recent argument by Brown and Rho
\cite{BR96} (and also \cite{gelmini})
has established that Walecka mean field theory is
equivalent to a chiral Lagrangian mean-field theory with the
parameters of the Lagrangian scaling a la Brown and Rho (``BR
scaling'') \cite{BR91}. In \cite{sbmr}, a Walecka-type Lagrangian
with BR-scaling parameters was constructed and shown to describe
the nuclear matter ground state as successfully as the effective
chiral action of Furnstahl et al.\ does. Such a Lagrangian has also
been shown to possess thermodynamic properties that are consistent
with Landau Fermi liquid structure of nuclear matter \cite{smr}.
This is the approach we shall use in this paper. Similar ideas were
developed by Gerry Brown but using different
arguments~\cite{gebhenley}.
\subsection{Fermi-liquid fixed points}
\indent\indent
A conceptually important point in our arguments is that Landau
Fermi-liquid theory~\cite{landau,migdal,noziere,baym-pethick} is an
effective field theory with fixed points~\cite{shankar}. In this
paper, we will not attempt to show that the Fermi-liquid theory for
nuclear matter is also a fixed-point field theory. We will simply
take the result established in \cite{shankar} and implement its
implications in our scheme.

One of the principal consequences of this identification is that
the nucleon effective mass (which will be referred to as ``Landau
effective mass") and Landau quasiparticle interactions (defined
below) are fixed-point quantities with vanishing $\beta$ functions.
Our goal is to connect these fixed-point quantities to BR scaling
parameters that figure in effective chiral Lagrangians appropriate
for dense medium. We are thereby combining two effective field
theories, chiral Lagrangian field theory and Landau Fermi-liquid
theory, into an effective field theory
 for dense matter in which BR scaling plays an
important role. We believe this ``marriage" is a successful one for
density at least up to that of nuclear matter. Going beyond that and
extrapolating into the regime of relativistic heavy-ion collisions are
guesses that need be verified a posteriori.

\subsection{A primer on Fermi-liquid theory}
\indent\indent
Before getting into our main calculation, we give a mini-primer on
Landau Fermi-liquid theory to define the quantities involved. We
should point out that once the fixed-point quantities are
identified in the chiral Lagrangian, then we can use all the
standard relations established in Landau's original theory.

Landau Fermi-liquid theory is a semi-phenomenological approach to
strongly interacting normal Fermi systems at small excitation
energies. It is assumed that a one-to-one correspondence exists
between the low-energy excitations of the Fermi liquid and that of
a non-interacting Fermi gas. The elementary excitations of the
Fermi liquid, which correspond to single particle degrees of
freedom of the Fermi gas, are called quasiparticles. The
quasiparticle properties, e.g.\ the mass, in general differ from
those of free particles due to interaction effects. In addition
there is a residual quasiparticle interaction, which is
parameterized in terms of the so called Landau parameters.

Fermi liquid theory is a prototype effective theory, which works
because there is a separation of scales. The theory is applicable
to low-energy phenomena, while the parameters of the theory are
determined by interactions at higher energies. The separation of
scales is due to the Pauli principle and the finite range of the
interaction.

Fermi-liquid theory has proven very useful \cite{baym-pethick} for
describing the properties of e.g.\ liquid $^3$He and provides a
theoretical foundation for the nuclear shell model~\cite{migdal}
as well as nuclear dynamics of low-energy excitations~\cite{speth,BWBS}.

The interaction between two quasiparticles ${\pb_1}$ and ${\pb_2}$
at the Fermi surface of symmetric nuclear matter can be written in
terms of a few spin and isospin invariants \cite{BSJ}
\be\label{qpint}
{ f}_{\pb_1\sigma_1\tau_1,\pb_2\sigma_2\tau_2}&=&
\frac{1}{N(0)}\left[F(\cos
\theta_{12})+F^\prime(\cos
\theta_{12})\taub_1\cdot
\taub_2+G(\cos \theta_{12})\sigmab_1\cdot
\sigmab_2\phantom{\frac{\qb^{\, 2}}{k_f^2}}\right.\nonumber\\&+&\left.
G^\prime(\cos \theta_{12})\sigmab_1\cdot
\sigmab_2\taub_1\cdot\taub_2
+\frac{\qb^{\, 2}}{k_f^2}H(\cos
\theta_{12})S_{12}(\qbhat)\right.\nonumber\\&+&\left.
\frac{\qb^{\, 2}}{k_f^2}H^\prime(\cos \theta_{12})S_{12}
(\qbhat)\taub_1\cdot\taub_2\right]
\ee
where $\theta_{12}$ is the angle between ${\pb_1}$ and ${\pb_2}$
and $N(0)=\lambda k_F m_N^\star/(2\pi^2)$ is the density of states
at the Fermi surface (we use natural units where $\hbar=1$ and denote
by $m_N^\star$ the (Landau) effective mass of the nucleon to be distinguished
from the BR-scaling mass $M_N^\star$). The
spin and isospin degeneracy factor $\lambda$ is equal to
4 in symmetric nuclear
matter. Furthermore, $\qb=\pb_1-\pb_2$ and
\bq\label{tensor}
S_{12}(\qbhat) = 3
\sigmab_1\cdot\qbhat\sigmab_2\cdot\qbhat -
\sigmab_1\cdot\sigmab_2,
\eq
where $\qbhat = \qb/\mid\qb\mid$. The tensor interactions $H$ and
$H^\prime$ are important for the axial charge, which we consider in
Section 4. The functions $F, F^\prime, \dots$ are expanded in
Legendre polynomials,
\bq
F(\cos \theta_{12})=\sum_\ell F_\ell P_\ell(\cos \theta_{12}),
\eq
with analogous expansions for the spin- and isospin-dependent
interactions. The energy of a quasiparticle\footnote{Below we omit
the spin and isospin indices $\sigma$ and $\tau$ from our formulas
to avoid overcrowding, except where needed to avoid ambiguities. We
will also omit the space and time dependence of the quantities,
e.g., $\epsilon\equiv\epsilon ({\Br },t)$ } with momentum
$p=|{\pb}|$, spin $\sigma$ and isospin $\tau$ is denoted by
$\epsilon_{p,\sigma,\tau}$ and the corresponding quasiparticle
number distribution by $n_{p,\sigma,\tau}$. The effective mass of a
quasiparticle on the Fermi surface is defined by
\bq
\left.\frac{d\epsilon_{p}}{dp}\right|_{p=\kF} = \frac{\kF}{m_N^\star}.
\eq
By using Galilean invariance one finds a relation between the
effective mass and the velocity dependence of the quasiparticle
interaction
\bq
\label{eff-mass}
\frac{m_N^\star}{m_N}=1 + \frac{F_1}{3}=\left(1-\frac{\tilde{F}_1}{3}
\right)^{-1},
\eq
where $\tilde{F}_l=(m_N/m_N^\star)F_l$, with analogous definitions
for $\tilde{F}_l^\prime$ etc..
\section{Electromagnetic Current}\label{em}
\indent\indent
We will first give a brief derivation of the Landau-Migdal formula
for the convection current for a particle of momentum $\Bk$
sitting on top of the Fermi sea responding to a slowly varying
electromagnetic (EM) field. We will then analyze it in terms of the
specific degrees
of freedom that contribute to the current. This will be followed by
a description in terms of a chiral Lagrangian as
discussed in \cite{FR96}. This procedure will provide the link between the
two approaches.
\subsection{Landau-Migdal formula for the convection current}\label{migdal}
\indent\indent
Following Landau's original reasoning adapted by Migdal to nuclear systems,
we start with
the convection current given by~\footnote{More precisely, this is a
matrix element of the current operator corresponding to the response
of a nucleon (proton
or neutron) sitting on top of the Fermi sea to the EM field. The sum over
spin and isospin and the momentum integral go over all occupied states up to
the valence particle.
What we want is a current operator and
it is deduced after the calculation is completed. One can of course
work directly with the operator but the result is the same.}
\be
\Jvec=\sum_{\sigma,\tau}\int \frac{d^3
p}{(2\pi)^3}(\Bnabla_p\epsilon_p) n_p\frac 12 (1+\tau_3)\label{landauJ}
\ee
where the sum goes over the spin $\sigma$ and isospin $\tau$ which
in spin- and isospin-saturated systems may be written as a trace
over the $\sigma$ and $\tau$ operators. We consider a variation of
the distribution function from that of an equilibrium state
\bq
n_p = n_p^0 + \delta n_p,
\eq
where the superscript $0$ refers to equilibrium. The variation of
the distribution function induces a variation of the quasiparticle
energy
\bq
\epsilon_p = \epsilon_p^0 +\delta\epsilon_p.
\eq
In the equilibrium state the current is zero by symmetry, so we
have
\be
\Jvec&=&\sum_{\sigma,\tau}\int \frac{d^3 p}{(2\pi)^3}
\left((\Bnabla_p\epsilon^0_p)\delta n_p+(\Bnabla_p\delta \epsilon_p)
n^0_p \right)\frac 12 (1+\tau_3),\nonumber\\
&=&\sum_{\sigma,\tau}\int \frac{d^3
p}{(2\pi)^3} \left((\Bnabla_p\epsilon^0_p)\delta n_p-(\Bnabla_p n^0_p)
\delta \epsilon_p) \right)\frac 12 (1+\tau_3)
\label{J}
\ee
to linear order in the variation. We consider a proton/neutron added at
the Fermi surface of a system in its ground state. Then
\be
\delta n_p=\frac{1}{V}\delta^3 (\pb-\bm{k})\frac{1\pm\tau_3}{2}\label{deltan}
\ee
and
\be
\Bnabla_p n^0_p=-\frac{\Bp}{k_F}\delta (p-k_F).
\ee
where $\bm{k}$ with $|\bm{k}|=k_F$ is the momentum of the
quasiparticle. The modification of the quasiparticle energies due
to the additional particle is given by
\be
\delta\epsilon_{p\sigma\tau}=\sum_{\sigma^\prime,\tau^\prime}
\int \frac{d^3 p^\prime}{(2\pi)^3}
f_{p\sigma\tau,p^\prime\sigma^\prime\tau^\prime}
\delta n_{p^\prime\sigma^\prime\tau^\prime}\label{deltaepsilon}.
\ee

Combining  (\ref{qpint}), (\ref{J}), (\ref{deltan}) and
(\ref{deltaepsilon}) one finds that the first term of (\ref{J})
gives the {\it operator}
\be
\Jvec^{(1)}=\frac{\bm{k}}{m_N^\star} \frac{1+\tau_3}{2},\label{qpJ}
\ee
where $\bm{k}$ is taken to be at the Fermi surface. The second term
yields
\be
\delta \Jvec=\delta \Jvec_s +\delta \Jvec_v=\frac{\bm{k}}{m_N}
\left(\frac{\tilde{F}_1
+\tilde{F}_1^\prime \tau_3}{6}\right)\label{deltaJ},
\ee
where
\be
\delta \Jvec_s &=&
 \frac{\Bk}{m_N^\star}\frac 12
\frac{F_1}{3},\label{deltas}\\
\delta \Jvec_v &=& \frac{\bm{k}}{m_N^\star} \frac{\tau_3}{2}\frac{F^\prime_1}{3}
= \frac{\bm{k}}{m_N^\star} \frac{\tau_3}{2} \frac{F_1}{3}
 +\frac{\bm{k}}{m_N^\star} \frac{\tau_3}{2}
 \frac{F^\prime_1-F_1}{3}.
\label{deltav}
\ee
Putting everything together we recover the well known result of
Migdal \cite{migdal,BWBS}
\bq
\Jvec=\frac{\bm{k}}{m_N}g_l = \frac{\bm{k}}{m_N} \left(\frac{1+\tau_3}{2}+\frac{1}{6}
(\tilde{F}^\prime_1-\tilde{F}_1)\tau_3\right),\label{Jtotal}
\eq
where
\bq
g_l=\frac{1+\tau_3}{2}+\delta g_l\label{gl}
\eq
is the orbital gyromagnetic ratio and
\bq
\delta g_l=\frac{1}{6}
(\tilde{F}^\prime_1-\tilde{F}_1)\tau_3.\label{deltagyro}
\eq
Thus, the renormalization of $g_l$ is purely isovector. This is due
to Galilean invariance, which implies a cancellation in the
isoscalar channel.

We have derived Migdal's result using standard Fermi-liquid theory
arguments. This result can also be obtained \cite{bentz} by using
the Ward identity, which follows from gauge invariance of the
electro-magnetic interaction. This is of course physically
equivalent to the above formulation. We shall now identify specific
hadronic contributions to the current (\ref{Jtotal}) in two ways: the
Fermi-liquid theory approach and the chiral Lagrangian approach.
\subsection{Pionic contribution}
\subsubsection{\it Fermi-liquid theory approach}
\indent\indent
In this approach, all we need to do is to compute the Landau parameter
$F_1$ from the pion exchange.
The one-pion-exchange contribution to the quasiparticle interaction
is \footnote{In a relativistic formulation sketched in Appendix A, we
can Fierz the one-pion exchange. Done in this way, the Fierzed
scalar channel is canceled by a part of the vector channel and the
remaining vector channel makes a natural contribution to the pionic
piece of $F_1$.}
\be
f^{\pi -exch.}_{\pb\sigma\tau,\pb^\prime\sigma^\prime\tau^\prime} =
-P_\sigma P_\tau V_\pi (q)= \frac 13
\frac{f^2}{m_\pi^2}
\frac{\qb^2}{\qb^2+m_\pi^2}\left(S_{12} (\qbhat)
+\frac 12 (3-\sigmab\cdot\sigmab^\prime)
\right)\frac{3-\taub\cdot\taub^\prime}{2}\label{vpi}
\ee
where $\qb=\pb-\pb^\prime$ and $f=g_{\pi NN} (m_\pi/2m_N)\approx 1$.
The one-pion-exchange contributions to the Landau parameters
relevant for the convection current are
\bq
\frac{F_1(\pi)}{3}= -F_1^\prime(\pi) = -\frac{3f^2 m_N^\star}{8\pi^2 k_F}
I_1\label{F1pi}
\eq
where
\be
I_1=\int_{-1}^1 dx \frac{x}{1-x+\frac{m_\pi^2}{2k_F^2}}= -2 +
(1+\frac{m_\pi^2}{2k_F^2})\ln (1+\frac{4k_F^2}{m_\pi^2}).\label{I1}
\ee
Thus, from Eq.(\ref{deltagyro}),
the one-pion-exchange contribution to the gyromagnetic ratio
is
\bq
\delta g_l^\pi=\frac{\mn}{\kF}\frac{f^2}{4\pi^2}I_1\tau_3.
\label{gyro-pi}
\eq
In Section 3.3,  we include
contributions also from other degrees of freedom.
\subsubsection{\it Chiral Lagrangian approach}
\indent\indent
In the absence of other meson degrees of freedom, we can
simply calculate Feynman diagrams given by a chiral Lagrangian
defined in the matter-free space. Nonperturbative effects due to
the presence of heavy mesons introduce a subtlety that will be
treated below.
\begin{figure}
\centerline{\epsfig{file=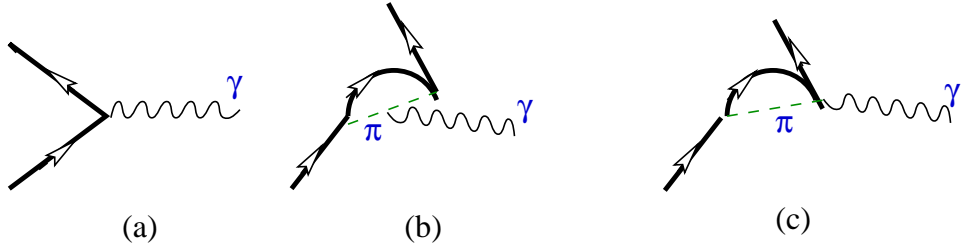,width=5in}}
\caption[f1]{\small \label{miyazawa}Feynman diagrams contributing
to the EM convection current in
effective chiral Lagrangian field theory. Figure (a) is the
single-particle term and (b, c) the next-to-leading chiral order
pion-exchange current term. Figure (c) does not contribute to the convection
current; it renormalizes the spin gyromagnetic ratio.}
\end{figure}

In the leading chiral order, there is the single-particle
contribution Figure \ref{miyazawa}a which for a particle on the Fermi surface
with the momentum $\Bk$ is given by
\be
\Jvec_{1-body}=\frac{\bm{k}}{m_N} \frac{1+\tau_3}{2}.\label{J1}
\ee
Note that the nucleon mass appearing in (\ref{J1}) is the
free-space mass $m_N$ as it appears in the Lagrangian, not the
effective mass $m_N^\star$ that enters in the Fermi-liquid approach,
(\ref{qpJ}).
To the next-to-leading order, we have
two ``soft-pion" terms as discussed in \cite{KDR,chemrho,pmr}. To
the convection current we need, only Fig. \ref{miyazawa}b
contributes\footnote{We should recall a well-known caveat here
discussed already in \cite{chemrho}. If one were to blindly
calculate the convection current coming from Fig. 1b, there would
be a gauge non-invariant term that is present because the hole line
is off-shell. Figure 1c contains also a gauge non-invariant term
which is exactly the same as in Figure 1b but with an opposite
sign, so in the sum of the two graphs, the two cancel exactly so
that only the gauge-invariant term survives. Of course we now know
that the off-shell dependence is not physical and could be removed
by field redefinition ab initio.},
\be
\Jvec_{2-body}=\frac{\bm{k}}{\kF}\frac{f^2}{4\pi^2} I_1\tau_3 =
\frac{\bm{k}}{\mn}\frac{1}{6}
(\tilde{F}^\prime_1(\pi)-\tilde{F}_1(\pi))\tau_3.
\label{J2}
\ee
We should emphasize that the Landau parameters $\tilde{F}_1$ and
$\tilde{F}_1^\prime$ are entirely fixed by chiral symmetry for any density..

The sum of (\ref{J1}) and (\ref{J2}) agrees precisely with the
Fermi-liquid theory result (\ref{Jtotal},\ref{F1pi},\ref{gyro-pi}).
This formula first derived in \cite{br87} in connection with the
Landau-Migdal parameter is of course the same as the Miyazawa
formula \cite{miyazawa} derived nearly half a century ago.  Note
the remarkable simplicity in the derivation starting from a chiral
Lagrangian. However, we should caution that there are some
non-trivial assumptions to go with the validity of the formula.
As we will see shortly, we will not have this luxury of simplicity when
other degrees of freedom enter.
\subsection{Vector-meson contributions and BR scaling}
\indent\indent
So far we have computed only the pion contribution to $g_l$. In
nuclear physics, more massive degrees of freedom such as the vector
mesons $\rho$ and $\omega$ of mass $700\sim 800$ MeV and the scalar
meson $\sigma$ of mass $600\sim 700$ MeV play an important role.
When integrated out from the chiral Lagrangian, they give rise to
effective four-Fermion interactions\footnote{For the moment, we
make no distinction as to whether one is taking into account BR
scaling or not. For the Fermi-liquid approach, this is not relevant
since the parameters are not calculated. However with chiral
Lagrangians, we will specify the scaling which is essential.}:
\be\label{four-Fermion}
{\cal L}_4=\frac{C_\phi^2}{2}(\bar{N}N)^2 -\frac{C_\omega^2}{2}
(\bar{N}\gamma_\mu N)^2 -\frac{C_\rho^2}{2}(\bar{N}\gamma_\mu\tau
N)^2 +\cdots
\ee
where the coefficients $C's$ can be identified with
\be\label{couplings}
C_M^2=\frac{g_M^2}{{m_M}^2} \ \ \ {\rm with}\ \ \ M=\phi,\
\rho,\ \ \omega.
\ee
Such interaction terms are ``irrelevant" in the renormalization
group flow sense but can make crucial contributions by becoming
``marginal" in some particular kinematic situation. A detailed
discussion of this point can be found in \cite{shankar}. The
effective four-Fermion interactions play a key role in stabilizing
the Fermi liquid state and leads to the fixed points for
the Landau parameters. (The other fixed-point quantity, i.e., the
effective mass, is put in by fiat to keep the density fixed.)
In the two-nucleon systems studied in
\cite{PKMR,pkmr-EFT}, they enter into the next-to-leading order term of the
potential, which is crucial in providing the cut-off independence
found for cut-off masses $\gsim m_\pi$.
\subsubsection{\it Fermi-liquid theory approach}
\indent\indent
Again it suffices to compute the Landau
parameters coming from the velocity-dependent part of
heavy meson exchanges.
We treat the effective four-Fermion interaction (\ref{four-Fermion})
in the Hartree approximation. Then the only velocity-dependent
contributions are due to the current couplings mediated by $\omega$
and $\rho$ exchanges. The corresponding contributions to the Landau
parameters are
\bq\label{f1-omega}
F_1(\omega)=-C_\omega^2\frac{2\kF^2}{\pi^2\mu}
\simeq-C_\omega^2\frac{2\kF^2}{\pi^2\mn}
\eq
and
\bq\label{f1p-rho}
F_1^\prime(\rho)=-C_\rho^2\frac{2\kF^2}
{\pi^2\mu}\simeq-C_\rho^2\frac{2\kF^2}{\pi^2\mn},
\eq
where $\mu$ is the baryon chemical potential and the final
expressions correspond to the non-relativistic limit.

Now the calculation of the convection current and the nucleon
effective mass with the interaction (\ref{four-Fermion}) in the Landau
method goes through the same way as in the case of the pion. The
net result is just Eq.(\ref{Jtotal}) including the
contribution of the contact interactions
(\ref{f1-omega},\ref{f1p-rho}), i.e.,
\be
\tilde{F}_1&=&\tilde{F}_1(\pi) +\tilde{F}_1(\omega),\\
\tilde{F}^\prime_1&=&\tilde{F}^\prime_1(\pi) +\tilde{F}^\prime_1(\rho).
\ee
Similarly, the nucleon effective mass is determined by
(\ref{eff-mass}) with
\be
F_1=F_1 (\pi) +F_1 (\omega).
\ee
\subsubsection{\it Chiral Lagrangian approach}
\indent\indent
The most efficient way to bring in the vector mesons into the
chiral Lagrangian is to implement BR scaling in the parameters of
the Lagrangian. We shall take the masses of the relevant degrees of
freedom to scale according to the BR
scaling~\cite{BR91}\footnote{In this paper, we are not addressing
how this relation was arrived at since our main objective is to
connect the scaling parameter $\Phi$
to many-body interactions and its link to the quark-antiquark
condensate in the medium-modified ``vacuum" does not
enter directly into our discussion. But it may be useful for the sake
of record to recall
that this relation was first written down using the Skyrme
Lagrangian embedded in medium with the scaling given by the
expectation value of the scalar that figures in the trace anomaly
of QCD~\cite{BR91}. Since this relation was first proposed, many
authors have attempted to ``derive" this scaling relation using
various QCD-motivated models as well as sum-rule-type arguments.
None of them has succeeded to
reproduce this relation. The reasons for this are
multi-fold but one of the main reasons is that the scalar field that
enters in the scaling has not been correctly identified. As argued
in \cite{sbmr}, the scalar field that dials BR scaling is the
``quarkonium" component of the trace anomaly, not the hard
``gluonic" component. The latter dominates the trace
anomaly but in the effective theory we are considering, this is
integrated out with its effects lodged in higher-dimensional
operators in the effective Lagrangian. In medium, as density is
increased and the chiral transition point is approached, the ``mended
symmetry" argument of Weinberg~\cite{mendedweinberg} as interpreted by Beane
and van Kolck~\cite{beane} suggests that the scalar contributing to
the trace anomaly that plays an important role in the scaling of
hadron properties is the scalar that makes up the
fourth component of $O(4)$ in linear sigma model. This structure
immediately gives, via a Nambu-Jona-Lasinio mechanism developed in
\cite{BBR}, the hadron scaling relation (\ref{BRscaling}). It has been
pointed out to us by Gerry Brown that this picture is supported by
a detailed lattice analysis of Liu et al.\ \cite{liu} for the source
of the mass of a constituent quark. Indeed most of the mass of the
light-quark hadron is shown to arise from the dynamical symmetry
breaking and hence is intricately tied to the change of the vacuum implied
in (\ref{BRscaling}).}
\be
\frac{M^\star_N}{m_N}\approx\frac{m_\omega^\star}{m_\omega}
\approx\frac{m_\rho^\star}{m_\rho}
\approx \frac{m_\phi^\star}{m_\phi}\approx \frac{f_\pi^\star}{f_\pi}
\equiv \Phi.\label{BRscaling}
\ee
Here $M_N^\star$ is a BR-scaling nucleon mass which will turn out
to be different from the Landau effective mass $m_N^\star$ \cite{FR96}. For
our purpose, it is more convenient to integrate out the vector and
scalar fields and employ the resulting four-Fermi interactions
(\ref{four-Fermion}). The coupling coefficients are modified compared
to Eq. (\ref{couplings}), because the meson masses
are replaced by effective ones:
\be
C_M^2=\frac{g_M^2}{{m_M^\star}^2} \ \ \ {\rm with}\ \ \ M=\phi,\
\rho,\ \ \omega.
\ee
The coupling constants may also scale~\cite{sbmr} but we omit
their density dependence for the moment.

\vskip 0.2cm
$\bullet$ {\it The relation between the BR factor $\Phi$ and
$F_1^\omega$}

\vskip 0.2cm
The first thing we need is the relation between the BR-scaling
factor $\Phi$ which was proposed in \cite{BR91} to reflect the
quark condensate {\it in the presence of matter} and the
contribution to the Landau parameter $F_1$ from the isoscalar
vector ($\omega$) meson. For this we first calculate the Landau effective
mass $m_N^\star$ in the presence of the pion and the $\omega$
fields~\cite{FR96}
\be
\frac{m_N^\star}{m_N}=1+\frac 13 (F_1 (\omega) +F_1 (\pi))=\left(1-\frac 13
(\tilde{F}_1 (\omega) +\tilde{F}_1 (\pi))\right)^{-1}.\label{mstar1}
\ee
Next we compute the nucleon self-energy
using the chiral Lagrangian. Given the single
quasiparticle energy $\epsilon_p$, we get the effective mass as in \cite{FR96}
\be
\frac{m_N^\star}{m_N}=\frac{k_F}{m_N} (\frac{d}{dp}\epsilon_p|_{p=k_F})^{-1}
=\left(\Phi^{-1} -\frac 13 \tilde{F}_1 (\pi)\right)^{-1}.\label{mstar2}
\ee
Comparing (\ref{mstar1}) and (\ref{mstar2}), we obtain the important
result
\be
\tilde{F}_1 (\omega)=3(1-\Phi^{-1}).\label{mainrelation}
\ee
This is an intriguing relation. It shows that the BR factor,
which was originally proposed as a precursor manifestation of the
chiral phase transition characterized by the vanishing of the quark
condensate at the critical point \cite{BR91}, is intimately related
(at least up to $\rho\approx \rho_0$)
to the Landau parameter $F_1$, which describes the quasiparticle
interaction in a particular channel.  We believe that the BR factor
can be computed by QCD sum-rule methods
or obtained from current algebra relations such as the
Gell-Mann-Oakes-Renner (GMOR) relation evaluated in-medium.
As was shown in \cite{FR96}, Eq. (\ref{mainrelation}) implies that
the BR factor governs in some, perhaps, intricate way
low-energy nuclear dynamics. This suggests a possible ``dual"
description at low density between what is given in QCD variables (e.g., quark
condensates) and what is given in hadronic variables (e.g., the
Landau parameter), somewhat reminiscent of the quark-hadron
duality in heavy-light-quark systems \cite{qhduality}.

\vskip 0.2cm
$\bullet$ {\it How to calculate the convection current in the
presence of BR scaling}

\vskip 0.2cm
In the presence of the BR scaling, a non-interacting nucleon in the chiral
Lagrangian propagates with a mass
$M_N^\star$, not the free-space mass $m_N$. Thus, the single-particle
current Fig. \ref{miyazawa}a is {\it not} given by (\ref{J1}) but
instead by
\be
\Jvec_{1-body}=\frac{\bm{k}}{M_N^\star} \frac{1+\tau_3}{2}.\label{onebody}
\ee
Now the current (\ref{onebody}) on its own does not carry conserved charge
as long as $M_N^\star\neq m_N$. This means that two-body currents
are indispensable to restore charge conservation. Note that the situation
is quite different from the case of Fermi-liquid theory. In the latter case,
the quasiparticle propagates with the Landau effective mass $m_N^\star$ and
it is the gauge invariance that restores $m_N^\star$ to $m_N$.\footnote{In
condensed matter physics, this is related to a phenomenon associated
with the cyclotron frequency which may be
referred to as  ``Kohn effect.''\cite{kohn}}
This clearly indicates that gauge invariance is more intricate when BR
scaling is implemented. Indeed
if the notion of BR scaling and the associated
chiral Lagrangian is to make sense, we have to recover charge
conservation from higher-order terms in the chiral Lagrangian. This
constitutes a strong constraint on the theory.

Let us now calculate the
contributions from the pion and heavy-meson degrees of freedom. The
pion contributes in the same way as before, so we can carry over
the previous result of Fig. \ref{miyazawa}b,
\be
\Jvec_{2-body}^{\pi}=\frac{\bm{k}}{m_N}\frac 16 (\tilde{F}_1^\prime (\pi)
-\tilde{F}_1 (\pi))\tau_3.\label{pi}
\ee
This is of the same form as (\ref{J2}) obtained in the {\it absence} of
BR scaling. It is in fact identical to (\ref{J2}) if we assume that
one-pion-exchange graph {\it does not scale} in medium at least up to nuclear
matter density. This assumption is supported by observations in pion-induced
processes in heavy nuclei~\footnote{In the early discussion of BR scaling
in \cite{BR91}, the mass parameter for an in-medium pion,
$m_\pi^\star$, in the effective chiral Lagrangian was
taken to scale down
as $\sim \sqrt{\Phi}$. However chiral perturbation theory in
medium predicts the ``pole mass" of the pion not to scale up to nuclear
matter density~\cite{thorsson}. In fact a recent analysis of deeply bound
pionic states in heavy nuclei~\cite{waas} shows that the pole mass of the pion
could be a few per cents higher than the free-space value at
nuclear matter density. The $m_\pi^\star$
in our in-medium effective chiral Lagrangian is not necessarily the pole
mass and so it is not clear how to implement this empirical information
into our theory. What we shall assume in this paper is that our $m_\pi^\star$
does not scale. This means that the observation that the one-pion-exchange
potential does not scale implies that the constant $g_A^\star/f_\pi^\star$
remains unscaling at least up to normal nuclear matter density. At high density
above normal nuclear matter density, however,
$g_A^\star$ will stabilize to 1 while
$f_\pi^\star$ will continue to drop and hence the coupling-constant ratio will
increase.}. In what follows, we will make this assumption implicitly.

\begin{figure}
\centerline{\epsfig{file=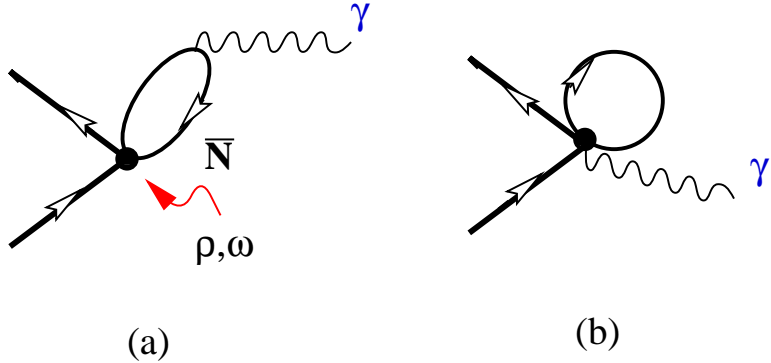,width=4in}}
\caption[f1]{\small (a) Feynman diagram contributing to the EM
 convection current from four-Fermi interactions corresponding to the
$\omega$ and $\rho$ channel (contact interaction indicated by the
blob) in effective chiral Lagrangian field theory. Th $\bar{N}$
denotes the anti-nucleon state that is given in the chiral
Lagrangian as a $1/m_N$ correction and the one without arrow is a
Pauli-blocked or occupied state. (b) The equivalent
graph in heavy-fermion formalism with the anti-nucleon line shrunk to
a point.}\label{nbar}
\end{figure}

The contributions from the vector-meson degrees of freedom are a bit trickier.
They are given by Fig. \ref{nbar}.
Both the $\omega$ (isoscalar) and $\rho$ (isovector) channels
contribute through the antiparticle intermediate state as shown in
Fig. \ref{nbar}a. The antiparticle is explicitly indicated in the
figure. However in the heavy-fermion formalism, the backward-going
antinucleon line should be shrunk to a point as Fig. \ref{nbar}b,
leaving behind an explicit $1/m_N$ dependence folded with a factor
of nuclear density signaling the $1/m_N$ correction in the chiral
expansion. One can interpret Fig. \ref{nbar}a as saturating the corresponding
counter term although this has to be yet verified by writing the
full set of counter terms at the same order.  These terms have been
evaluated in \cite{FR96} with Fig. \ref{nbar}a
\be
\Jvec_{2-body}^{\omega}&=&\frac{\Bk}{m_N}\frac 16 \tilde{F}_1 (\omega),
\label{omega}\\
\Jvec_{2-body}^{\rho}&=&\frac{\Bk}{m_N}\frac 16 \tilde{F}_1^\prime
(\rho)\tau_3,\label{rho}
\ee
where $\tilde{F}_1 (\omega)$ and $\tilde{F}_1^\prime(\rho)$ are given
by Eqs. (\ref{f1-omega},\ref{f1p-rho}) with $m_N$ replaced by $M_N^\star$.
The total current given by the sum of (\ref{onebody}), (\ref{pi}),
(\ref{omega}) and (\ref{rho}) precisely agrees with the Fermi-liquid theory
result (\ref{Jtotal}) when we take
\be
\tilde{F}_1&=&\tilde{F}_1 (\omega)+\tilde{F}_1 (\pi),\\
\tilde{F}_1^\prime &=&\tilde{F}_1^\prime (\rho)+\tilde{F}_1^\prime (\pi).
\ee

The way in which this precise agreement comes about is nontrivial. What
happens is that part of
the $\omega$ channel restores the BR-scaled mass $M_N^\star$ back
to the free-space mass $m_N$ in the isoscalar current. (It has been known
since sometime that something similar
happens in the standard Walecka model (without pions and BR scaling)
\cite{kurasawa}). Thus, the
leading single-particle operator combines with the sub-leading
four-Fermi interaction to restore the charge conservation as
required by the Ward identity. This is essentially the ``back-flow
mechanism" which is an important ingredient in Fermi-liquid theory.
We describe below the
standard back-flow mechanism as given in textbooks \cite{pines}, adapted to
nuclear systems with isospin degrees of freedom, and
elucidate the connection to the results obtained with the chiral
Lagrangian in this section.
\subsubsection{\it The $\omega/q\rightarrow 0$ limit and
the ``back-flow current"}
\begin{figure}
\centerline{\epsfig{file=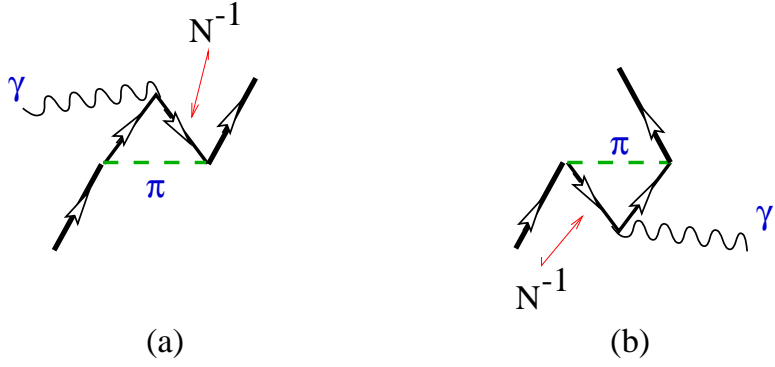,width=4in}}
\caption[f3]{\small Particle-hole contributions to the convection current.
Here backward-going nucleon line $N^{-1}$ denotes a hole. These graphs
vanish in the $q/\omega\rightarrow 0$ limit.}\label{ph}
\end{figure}

\indent\indent
The current so constructed is valid for a process occurring very
near the Fermi surface corresponding to the limit
$(\omega,\Bq)\rightarrow (0,\bf {0})$ where $q$ is the spatial
momentum transfer and $\omega$ is the energy transfer. In the diagrams
considered so far (Fig. \ref{miyazawa} and \ref{nbar}) the order of
the limiting processes does not matter. However, the particle-hole
contribution, which we illustrate in Fig. \ref{ph} with the pion
contribution\footnote{The relation we derive below holds in general regardless
of what is being exchanged as long as the exchanged particle has the right
quantum numbers.}, does depend on the order in which $q=|\Bq|$ and $\omega$
approach zero. Thus, in the limit
$q/\omega\rightarrow 0$, the particle-hole contributions vanish
whereas in the opposite case $\omega/q\rightarrow 0$, they do not.
This can be seen by examining the particle-hole propagator
\be
\frac{n_k (1-n_{k+q})}{q_0 +\epsilon_k-\epsilon_{k+q}+i\delta} -
\frac{n_{k+q} (1-n_k)}{q_0+\epsilon_k-\epsilon_{k+q}-i\delta}
\ee
where $(q_0, \Bq )$ is the four-momentum of the external (EM)
field. This vanishes if we set $q\rightarrow 0$ with $q_0$ non-zero
but its real part is non-zero if we interchange the limiting process
since for $q_0=0$ we have
\be
\frac{\Bq\cdot\kbhat }{-\Bq\cdot\Bk/m_N}\delta (k_F-k).
\ee
In the limit $\omega/q\rightarrow 0$, the particle-hole
contribution to the current is\footnote{See Appendix B for a brief derivation
of this expression with one pion exchange.}
\be
\Jvec_{ph}= - \frac{\Bk}{m_N} \left(\frac{\tilde{F}_1
+\tilde{F}_1^\prime \tau_3}{6}\right).\label{parthole}
\ee
Adding the particle-hole contribution (\ref{parthole})
to the Fermi-liquid result (\ref{Jtotal}) we obtain the current of a
{\it dressed} or localized quasiparticle
\be
\Jvec_{locQP}=\frac{\Bk}{m_N^\star} \left(\frac{1+\tau_3}{2}\right).
\label{locQP}
\ee
Note that $\Jvec_{ph}$ precisely cancels $\delta\Jvec$, Eq.(\ref{deltaJ}).
The current $\Jvec_{locQP}$ is the total current carried by the wave
packet of a localized quasiparticle with group velocity
${\bf {v}}_F=\frac{\Bk}{m_N^\star}$.
However, the physical situation corresponds to homogeneous (plane wave)
quasiparticle excitations. The current carried by a localized
quasiparticle equals that of a homogeneous quasiparticle
excitation modified by the so called back-flow current \cite{pines}.
The back-flow contribution $(\Jvec_{locQP}-\Jvec_{LM})$ is
just the particle-hole polarization current in the $\omega/q\rightarrow 0$
limit, Eq.(\ref{parthole}).
\section {Axial Charge Transitions}\label{axial}
\indent\indent
No one has yet derived the analogue to (\ref{Jtotal}) for the axial
current. Attempts using axial Ward identities in analogy to the
electromagnetic case have not met with success \cite{bentzaxial}.
The difficulty has presumably to do with the role of the Goldstone
bosons in nuclear matter which is not well understood. In this
section, we analyze the expression for the axial charge operator
obtained by a straightforward application of the Fermi-liquid theory
arguments of Landau and Migdal and compare this expression with
that obtained directly from the chiral Lagrangian using current
algebra. For the vector current we found precise agreement between
the two approaches.
\subsection{Applying Landau quasiparticle argument}
\indent\indent

The obvious thing to do is to simply mimic the steps used for the
vector current to deduce a ``Landau-Migdal" expression for the
axial charge operator. We use both methods developed above and find
that they give the same result.

In free space, the axial charge operator nonrelativistically is
$\sim
\Bsigma\cdot{\bf {v}}$ where ${\bf {v}}=\Bk/m_N$ is the
velocity. In the infinite momentum frame, it is the relativistic
invariant helicity ${\Bsigma}\cdot \nubhat$. It is thus tempting to
assume that near the Fermi surface, the axial charge operator for a
{\it local} quasiparticle in a wave packet moving with the group
velocity ${\bf v}_F=\Bk/m_N^\star$ is simply $\sim
{\Bsigma}\cdot{\bf v}_F$. This suggests that we take the axial
charge operator for a {\em localized} quasiparticle to have the
form
\be
 {\A0}^i_{locQP}= g_A\frac{\Bsigma\cdot\Bk}{m_N^\star}
\frac{\tau^i}{2}.
\label{totalcharge}
\ee
As in the vector current case, we take (\ref{totalcharge}) to be
the $\omega/q\rightarrow 0$ limit of the axial charge operator. The
next step is to compute the particle-hole contribution to Fig.3
(with the vector current replaced by the axial current) in the
$\omega/q\rightarrow 0$ limit. A simple calculation gives
\be
{\A0}^i_{ph}=-g_A\frac{\Bsigma\cdot\Bk}{m_N^\star}\frac{\tau^i}{2}
\Delta^\prime\label{phA0}
\ee
with
\be
\Delta^\prime=\frac{f^2k_F m_N^\star}{4m_\pi^2 \pi^2} (I_0-I_1)
\ee
where $I_1$ was defined in (\ref{I1}) and
\be
I_0=\int_{-1}^1 dx\frac{1}{1-x+\frac{m_\pi^2}{2k_F^2}}=
\ln\left(1+\frac{4k_F^2}{m_\pi^2}\right).
\ee
In an exact parallel to the procedure used for the vector current,
we take the difference
\be
{\A0}^i_{locQP}-{\A0}^i_{ph}
\ee
and identify it with the corresponding ``Landau axial charge'' (LAC):
\be
{\A0}^i_{LAC}=
g_A\frac{\Bsigma\cdot\Bk}{m_N^\star}\frac{\tau^i}{2}(1+\Delta^\prime).
\label{LAC}
\ee

Let us now 
rederive (\ref{LAC}) with an
argument analogous to that proven to be powerful for the convection
current. We shall do the calculation using the pion exchange only
but the argument goes through when the contact interaction
(\ref{four-Fermion}) is included. We begin by assuming that the
axial charge -- in analogy to (\ref{landauJ}) for the convection
current --  takes the form,
\be
{\A0}^i=g_A \sum_{\sigma\tau}\int \frac{d^3 p}{(2\pi)^3}
\Bsigma\cdot (\Bnabla_p\epsilon_p) n_p \frac{\tau^i}{2}\label{a0me}
\ee
where $n_p$ and $\epsilon_p$ are $2\times 2$ matrices with matrix
elements
\be
[n_p (\Br,t)]_{\alpha\alpha^\prime}=n_p
(\Br,t)\delta_{\alpha\alpha^\prime} +\Bs_p
(\Br,t)\cdot\Bsigma_{\alpha\alpha^\prime},
\ee
and
\be
[\epsilon_p (\Br,t)]_{\alpha\alpha^\prime} = \epsilon_p
(\Br,t)
\delta_{\alpha\alpha^\prime} +{\Beta}_p (\Br,t)\cdot
\Bsigma_{\alpha\alpha^\prime}
\ee
with
\be
{\Bs}_p (\Br,t)=\frac 12 \sum_{\alpha\alpha^\prime}
\Bsigma_{\alpha\alpha^\prime} [n_p (\Br,t)]_{\alpha^\prime\alpha}.
\label{sigmap}
\ee
In general $n=4$ in the spin-isospin space. But without loss of
generality, we could confine ourselves to $n=2$ in the spin space
with the isospin operator explicited as in Eq.(\ref{a0me}). Then
upon linearizing, we obtain from (\ref{a0me})
\be
{\A0}^i=g_A \sum_{\sigma\tau}\int \frac{d^3 p}{(2\pi)^3}
\left(\Bsigma\cdot(\Bnabla_p\epsilon^0_p)\delta n_{p\sigma\tau} -
\Bsigma\cdot (\Bnabla_p n^0_p)\delta\epsilon_{p\sigma\tau}\right)
\frac{\tau^i}{2}+\cdots\label{a0mes}
\ee
where
\be
\delta n_{p\sigma\tau}=\frac 1V \delta^3 (\Bp-\Bk )
\frac{1+\sigma_3}{2}\frac{\tau^i}{2}
\label{deltanp}
\ee
and
\be
\delta\epsilon_{p\sigma\tau}=\sum_{\sigma^\prime,\tau^\prime}
\int \frac{d^3 p^\prime}{(2\pi)^3}
f_{p\sigma\tau,p^\prime\sigma^\prime\tau^\prime}
\delta n_{p^\prime\sigma^\prime\tau^\prime}.\label{deltaen}
\ee
in analogy with (\ref{deltaepsilon}). The equation (\ref{a0mes}) is
justified if the density of polarized spins is much less than the
total density of particles (assumed to hold here).
The first term of (\ref{a0mes}) with (\ref{deltanp}) yields the
quasiparticle charge operator
\be
{\A0}^i_{QP}=g_A\frac{\Bsigma\cdot\Bk}{m_N^\star}\frac{\tau^i}{2}\label{A0}
\ee
while the second term represents the polarization of the medium,
due to the pion-exchange interaction (\ref{vpi})
\be
\delta {\A0}^i=g_A\frac{\Bsigma\cdot\Bk}{m_N^\star}\frac{\tau^i}{2}
\Delta^\prime.\label{deltaA0}
\ee
The sum of (\ref{A0}) and (\ref{deltaA0}) agrees precisely with the Landau
charge (\ref{LAC}).

It is not difficult to take into account the full Landau-Migdal
interactions (\ref{qpint}) which includes the one-pion-exchange
interaction as well as other contributions to the quasiparticle
interaction.
Thus, the general expression is obtained by making the replacement
\be\label{zero-minus}
\Delta^\prime \rightarrow \frac 13 G_1^\prime - \frac{10}{3}H_0^\prime
+\frac 43 H_1^\prime - \frac{2}{15}H_2^\prime
\ee
in (\ref{deltaA0}). This combination of Fermi-liquid parameters
corresponds to a $\ell = \ell ^\prime = 1, J=0$ distortion of the
Fermi sea \cite{BSJ}. We will see later that the result obtained
with the naive Landau argument may not be the whole story, since
the one-pion-exchange contribution disagrees, though by a small
amount, with the chiral Lagrangian prediction derived below.

\subsection{Chiral Lagrangian prediction}
\indent\indent
We now calculate the axial charge using our chiral Lagrangian that
reproduced the Landau-Migdal formula for the convection current.
Consider first only the pion-exchange contribution.
In this case we can take the unperturbed nucleon
propagator to carry the free-space mass $m_N$. The single-particle
transition operator corresponding to Fig.1a is given by
\be
{\A0}^i_{1-body}=g_A\frac{\Bsigma\cdot\Bk}{m_N}\frac{\tau^i}{2}.
\label{a01bod}
\ee
There is no contribution of the type of
Fig.1b because of the (G-)parity conservation.
The only contribution to the two-body current comes from
Fig.1c and is of the form \cite{delorme}
\be
{\A0}^i_{2-body}=g_A\frac{\Bsigma\cdot\Bk}{m_N}\frac{\tau^i}{2}\Delta
\label{a02bod}
\ee
with
\be
\Delta=\frac{f^2 k_F m_N}{2g_A^2m_\pi^2\pi^2}\left(I_0-I_1-
\frac{m_\pi^2}{2k_F^2}I_1\right).\label{Delta}
\ee
The factor $(1/g_A^2)$ in (\ref{Delta}) arose from replacing
$\frac{1}{f_\pi^2}$ by $\frac{g_{\pi NN}^2}{g_A^2 m_N^2}$ using the
Goldberger-Treiman relation.

Now consider what happens when  the vector degrees of freedom
are taken into account.  Within the approximation adopted, the only thing that
needs be done is to implement the BR scaling. The direct
intervention of the vector mesons $\rho$ and $\omega$ in the
axial-charge operator is suppressed by the chiral counting, so they
will be ignored here. This means that in the single-particle charge
operator, all that one has to do is to replace $m_N$ by
$M_N^\star=m_N\Phi$ in (\ref{a01bod}):
\be
{\A0}^i_{1-body}=g_A\frac{\Bsigma\cdot\Bk}{m_N\Phi}\frac{\tau^i}{2}
\label{a01bod*}
\ee
and that in the two-body charge operator (\ref{a02bod}), $f_\pi$
should be replaced by $f_\pi\Phi$ and $m_N$ by
$m_N\Phi$:
\be
{\A0}^i_{2-body}=g_A\frac{\Bsigma\cdot\Bk}{m_N\Phi}\frac{\tau^i}{2}
\Delta.\label{a02bod*}
\ee
In the two-body operator, there is a factor $(g_A/f_\pi)$ coming
from the $\pi NN$ vertex which as mentioned before,
is assumed to be non-scaling at
least up to nuclear matter density \cite{gebhenley,FR96}, in
consistency with the observation that the pion-exchange current
 does not scale in medium.

The total predicted by the chiral Lagrangian (modulo higher-order corrections)
is then
\be
g_A\frac{\Bsigma\cdot\Bk}{m_N\Phi}\frac{\tau^i}{2}
(1+\Delta).
\ee
which differs from the charge operator obtained by the Landau method,
(\ref{LAC}).
\section{Comparison Between Vector and Axial Currents}\label{comparison}
\subsection{Fermi-liquid theory vs. current algebra}
\indent\indent
An immediate question (to which we have no convincing answer) is whether or
not the difference between the
two approaches -- the Fermi-liquid vs. the chiral Lagrangian -- is genuine
or a defect in either or both of the approaches.
One possible cause of the difference could
be that {\it both} the assumed localized quasiparticle charge,
Eq.(\ref{totalcharge}),
{\it and} the effective axial charge, Eq.(\ref{a0me}), are incomplete.
We have looked for possible additional terms that could contribute but we have
been unable to find them.
So while not ruling out this possibility, we turn to the possibility
that the difference is genuine.

It is a well-known fact that the conservation of the vector current
assures that the electromagnetic charge or the weak vector charge
is $g_V=1$ but the conservation of the axial vector charge does not
constrain the value of the axial charge $g_A$ , that is, $g_A$ can
be anything. This is because the axial symmetry is spontaneously
broken. In the Wigner phase in which the axial symmetry would be
restored, one would expect that $g_A=1$. It therefore seems that
the Goldstone structure of the ``vacuum" of the nuclear matter is
at the origin of the difference.

To see whether there can be basic differences, let us look at the effect of
the pion field. The cancellation between the two-body current
$\Jvec_{2-body}^\pi$ (\ref{pi}) and $\Jvec_{ph}^\pi$
(\ref{parthole}) leaving only a term that changes $M_N^\star$ to
$m_N^\star$ in the one-body operator with a BR-scaling mass,
Eq.(\ref{onebody}), in the EM case can be understood as follows.
Both terms involve the two-body interaction mediated by a
pion-exchange. It is obvious how this is so in the latter.  To see
it in the former, we note that it involves the insertion of an EM
current in the propagator of the pion. Thus the sum of the two
terms corresponds to the insertion of an EM current in {\it all
internal hadronic lines} of the one-pion exchange self-energy graph
of the nucleon.  The two-body pionic current -- that together with
the single-particle current preserves gauge invariance -- is in turn
related to the one-pion-exchange potential
$V_\pi$. Therefore what is
calculated is essentially an effect of a nuclear force. Now
the point is that the density-dependent part of the sum (that is,
the ones containing one hole line) -- apart from a term that
changes $M_N^\star$ to $m_N^\star$ in (\ref{onebody}) -- vanishes
in the $\omega/q\rightarrow 0$ limit. In contrast, the cancellation
between (\ref{deltaA0}) and (\ref{phA0}) in the case of the axial charge,
has no corresponding
interpretation. While the one-pion-exchange interaction is involved
in the particle-hole term (\ref{phA0}),  (\ref{deltaA0}) cannot be
interpreted as an insertion of the axial vector current into the
pion propagator since such an insertion is forbidden by parity. In
other words, Eq.(\ref{deltaA0}) does not have a corresponding
Feynman graph which can be linked to a potential. {\it We interpret
this as indicating that there is no corresponding Landau formula
for the axial charge in the same sense as in the vector current
case.}

In a chiral Lagrangian formalism, each term is associated with a
Feynman diagram. As mentioned, there is no contribution to the
convection current from a diagram of the type Fig.1c (apart from a
gauge non-invariant off-shell term which cancels the counter part
in Fig.1b). Instead this diagram renormalizes the spin gyromagnetic
ratio. In contrast, the corresponding diagram for the axial current
{\it does} contribute to the axial charge (\ref{a02bod}). As first
shown in \cite{KDR}, the contribution from Fig.1c for both the
vector current and the axial-vector current is current algebra in
origin and constrained by chiral symmetry. Furthermore it does not
have a simple connection to nuclear force. While the convection
current is completely constrained by gauge invariance of the
electromagnetic field, and hence chiral invariance has little to
say,  both the EM spin current and the axial charge are principally
dictated by the chiral symmetry. {\it This again suggests that the
Landau approach to the axial charge cannot give the complete answer
even at the level of quasiparticle description.} There is however a
caveat here: in the Landau approach, the nonlocal pionic and local
four-Fermion interactions (\ref{four-Fermion}) enter together in an
intricate way as we saw in the electromagnetic case. Perhaps this
is also the case in the axial charge, with an added subtlety due to
the presence of Goldstone pions. It is possible that the difference
is due to the contribution of the four-Fermion interaction term to
(\ref{zero-minus}) which cancels out in the limit
$\omega/q\rightarrow 0$ but contributes in the $q/\omega\rightarrow
0$ limit. This term cannot be given a simple interpretation in
terms of chiral Lagrangians.
Amusingly the difference between the results (see below) turns out
to be small.
\subsection{Numerical comparison}
\indent\indent
To compare the two results, we rewrite the sum of (\ref{A0}) and
(\ref{deltaA0}), i.e., ``Landau axial charge" (LAC), using
(\ref{eff-mass}) and (\ref{F1pi})
\be
 {\A0}^i_{LAC}=g_A\frac{\Bsigma\cdot\Bk}{m_N\Phi}\frac{\tau^i}{2}(1+\tilde{\Delta})
\label{LACp}
\ee
where
\be
\tilde{\Delta}=\frac{f^2 k_F m_N\Phi}{4\pi^2 m_\pi^2} \left(I_0-I_1 +
\frac{3m_\pi^2}{2k_F^2}I_1\Phi^{-1}\right)\label{delta"}
\ee
and the sum of (\ref{a01bod*}) and (\ref{a02bod*}), i.e.,
the  ``current-algebra axial charge" (CAAC), as
\be
{\A0}^i_{CAAC}=g_A\frac{\Bsigma\cdot\Bk}{m_N\Phi}\frac{\tau^i}{2}
(1+\Delta)\label{CAAC}
\ee
where
\be
\Delta=\frac{f^2 k_F m_N}{2g_A^2m_\pi^2\pi^2}\left(I_0-I_1-
\frac{m_\pi^2}{2k_F^2}I_1\right).
\ee
We shall compare $\tilde{\Delta}$ and $\Delta$ for two densities
$\rho=\frac 12\rho_0$ ($k_F=1.50 m_\pi$) and $\rho=\rho_0$
($k_F=1.89m_\pi$) where $\rho_0$ is the normal nuclear matter
density $0.16/{\rm fm}^3$.

For numerical estimates, we take
\be
\Phi (\rho)=\left(1+0.28\frac{\rho}{\rho_0}\right)^{-1}
\ee
which gives $\Phi(\rho_0)=0.78$ found in QCD sum-rule calculations
\cite{sbmr,FR96}. Somewhat surprisingly, the resulting values for
$\tilde{\Delta}$ and $\Delta$ are close; they agree within 10\%.
For instance at $\rho\approx
\rho_0/2$, $\tilde{\Delta}\approx 0.48$ while $\Delta\approx 0.43$ and at
$\rho\approx \rho_0$, $\tilde{\Delta}\approx 0.56$ while
$\Delta\approx 0.61$. Whether this close agreement is coincidental
or has a deep origin is not known.
\section{Kaonic Fluctuation and Nuclear Matter}\label{kaon}
\indent\indent
The scaling relations we have established so far involve the BR-scaling
nucleon mass $M^\star$, the Landau effective mass $m_N^\star$ and the
in-medium constant $f_\pi^\star$. The connection to the scaling of the meson
masses was made indirectly through (\ref{four-Fermion}) that resulted from
integrating out heavy degrees of freedom of the appropriate quantum numbers.

As discussed in \cite{BR96,sbmr}, somewhat more direct information
on the link to the meson-mass scaling can be gotten from
kaon-nuclear interactions given in the tree approximations of the
effective chiral Lagrangian with BR scaling. In terms of
constituent quarks, the interaction of a kaon with a nucleon in a
medium -- since a kaon carries only one non-strange quark -- can be
related to one-third of nucleon-nucleon interaction.  This gives an
attraction of $\sim 190$ MeV in $K^-$-nuclear potential in
agreement with the kaonic atom data \cite{friedmangalmares}, i.e.,
$185\pm15$ MeV. Described in terms of scaling parameters, this
attraction follows immediately from BR scaling of the vector and
scalar masses as argued in \cite{BR96,sbmr}. This implies that
nuclear matter should be describable in terms of the same
Lagrangian in the non-strange sector. It was shown  in \cite{sbmr}
that with the scaling masses -- and with a similar scaling in the
vector coupling, all nuclear matter properties including a low
compression modulus $K\lsim 300$ MeV come out correctly. Since the
mean-field Lagrangian that figures here is in form equivalent to
Walecka's linear model with {\it all} parameters suitably scaling,
all the thermodynamical properties that are satisfied by the
Walecka model are also satisfied by the BR-scaling Lagrangian. This
connection thus offers 
a natural
interpretation in terms of Landau-Migdal Fermi liquid theory along
the line that Matsui developed for the Walecka model \cite{matsui}.
This provides us another support for the relation between BR
scaling and Landau Fermi-liquid fixed point parameters.
\section{Phenomenological Tests}\label{test}
\indent\indent
Most of the tests of the formulas derived in this paper have been
discussed elsewhere~\cite{sbmr,FR96,tests} but we present a few
crucial (and some new) ones here to make this paper self-contained.
\subsection{Landau effective mass}
\indent\indent
It is not obvious that the effective nucleon mass computed in the
chiral Lagrangian approach is directly connected to a measurable
quantity although quasielastic electron scattering from nuclei does
probe some kind of effective nucleon mass and Walecka model
describes such a process in terms of an effective mass. To the
extent that the bulk of $m_N^\star$ is related to the condensate
through BR scaling as we can see in (\ref{mstar2}),
the effective mass in the chiral Lagrangian
can be related to the quantity calculated in QCD sum-rule approach
for in-medium hadron masses. In BR scaling, the parameter $\Phi$ is
related to the scaling of the vector meson ($\rho$) mass. There are
several QCD sum-rule calculations for the $\rho$ meson in-medium
mass starting with \cite{SHLee}. The most recent one which closely
agrees with the Gell-Mann-Oakes-Renner formula in medium for the
pion decay constant $f_\pi^\star$ (see Eq.(\ref{BRscaling})) is the
one by Jin et al.\ \cite{jin}:
\be
\Phi(\rho_0)=0.78\pm 0.08.\label{Jinvalue}
\ee
We shall take this value in what follows but one should be aware of
the possibility that this value 
is not
quite firm\footnote{A caveat to this result was recently discussed
by Klingl et al.~\cite{klingl} who show that the QCD sum rule can
be saturated without the mass shifting downward by increasing the
vector meson width in medium. For a discussion of the empirical
constraints on the in-medium widths of vector mesons, see
Friman~\cite{friman97}.}. Given this, we can compute $m_N^\star$
using (\ref{mstar2}) for nuclear matter density since the pionic
contribution $\tilde{F}_1 (\pi)$ is known.
One
finds \cite{FR96}
\be
\frac{m_N^\star}{m_N}(\rho=\rho_0)\approx 0.70.\label{m*pred}
\ee
This can be tested in an indirect way by looking at certain
magnetic response of nuclei as described below. An additional
evidence comes from QCD sum-rule calculations. Again there are
caveats in the QCD sum-rule calculation for the nucleon mass even
in free-space and certainly more so in medium.
Nevertheless
the most recent result by Furnstahl
et al.~\cite{furnstahljin} is rather close to the prediction
(\ref{m*pred}):
\be
\left(\frac{m_N^\star (\rho_0)}{m_N}\right)_{QCD}=0.69^{+0.14}_{-0.07}.
\ee
\subsection{Orbital gyromagnetic ratio}
\indent\indent
If one writes the gyromagnetic ratio $g_l$ as
\be
g_l=\frac{1+\tau_3}{2} +\delta g_l
\ee
then the chiral Lagrangian prediction is
\be
\delta g_l=\frac 16 (\tilde{F}_1^\prime -\tilde{F}_1)\tau_3
=\frac 49 \left[\Phi^{-1} -1- \frac 12 \tilde{F}_1 (\pi)\right]\tau_3.
\label{chptd}
\ee
In writing the second equality we have used (\ref{F1pi}), (\ref{mainrelation})
and the nonet relation $\tilde{F}^\prime (\rho)=\tilde{F} (\omega)/9$.
At nuclear matter density, we get, using (\ref{Jinvalue}),
\be
\delta g_l (\rho_0)\approx 0.23\tau_3.
\ee
This agrees with the value extracted from a dipole sum rule
in $^{209}$Bi~\cite{schumacher},
\be
\delta g_l^{proton}=0.23\pm 0.03\label{experiment}
\ee
and agrees roughly with magnetic moment data in heavy
nuclei\footnote{Nuclear magnetic moments are complicated due to
conventional nuclear effects. To make a meaningful comparison, one
would have to extract all ``trivial" nuclear effects and this
operation brings in inestimable uncertainties.}. It should be
stressed that the gyromagnetic ratio provides a test for the
scaling nucleon mass at $\rho\approx
\rho_0$. It also gives a check of the relation
between the {\it baryon property} on the left-hand side of
Eq.(\ref{mainrelation}) and the {\it meson property} on the
right-hand side. Instead of using (\ref{Jinvalue}) as an input
to calculate $\delta g_l$, we could take the experimental value
(\ref{experiment}) to determine, using (\ref{chptd}), the BR-scaling 
factor $\Phi$ at $\rho\approx \rho_0$. We would of course
get (\ref{Jinvalue}), a value which is consistent with what one obtains
in the QCD sum-rule calculation and also in the in-medium GMOR relation.
\subsection{Axial charge transitions in heavy nuclei}
\indent\indent
The axial charge transition in heavy nuclei
\be
A(J^+)\leftrightarrow B(J^-)
\ee
with change of one unit of isospin $\Delta T=1$ provides a test of
the axial charge operator (\ref{CAAC}) or (\ref{LACp}). To check this,
consider the Warburton ratio $\epsilon_{MEC}$ \cite{warburton}
\be
\epsilon_{MEC}=M_{exp}/M_{sp}
\ee
where $M_{exp}$ is the {\it measured} matrix element for the axial
charge transition and $M_{sp}$ is the {\it theoretical}
single-particle matrix element. There are theoretical uncertainties
in defining the latter, so the ratio is not an unambiguous object
but what is significant is Warburton's observation that in heavy
nuclei, $\epsilon_{MEC}$ can be as large as 2:
\be
\epsilon_{MEC}^{HeavyNuclei}=1.9\sim 2.0.
\ee
More recent measurements -- and their analyses -- in different
nuclei~\cite{otherexp}
quantitatively confirm this result of Warburton.

To compare our theoretical prediction with the Warburton analysis,
we calculate the same ratio using (\ref{CAAC})\footnote{This formula differs
from what was obtained in \cite{kuboderarho} in that here the non-scaling in
medium of the
pion mass and the ratio $g_A/f_\pi$ is taken into account. We believe that
the scaling used in \cite{kuboderarho} (which amounted to having $\Delta/\Phi$
in place of $\Delta$ in (\ref{epsilonth})) is not correct.}
\be
\epsilon_{MEC}^{CAAC}=\Phi^{-1} (1+\Delta).\label{epsilonth}
\ee
The enhancement corresponding to the ``Landau formula"
(\ref{LACp}) is obtained by replacing $\Delta$ by
$\tilde{\Delta}$ in (\ref{epsilonth}). Using the value for $\Phi$
and $\Delta$ at nuclear matter density, we find\footnote{An
accurate measurement of the factor $\epsilon_{MEC}$ has been made
in the $A=12$ system by Minamisono et al.\ \cite{minamisono}. The
presently available data indicate that the experimental value is
quite high, $\sim 1.6$. Just to have an idea how the present theory
predicts, let us assume that the average density appropriate for
the process in the $A=12$ system is $\rho\approx \rho_0/2$ for
which we found above $\tilde{\Delta}
\approx 0.48$ (or $\Delta\approx 0.43$) and $\Phi (\rho_0/2)^{-1}\approx 1.14$.
Thus the prediction for this system
is that $\epsilon_{MEC}\approx 1.63\sim 1.69$. This is
consistent with what Minamisono et al.\ found. To make a direct comparison
with the experimental data of \cite{minamisono}, we would have to take into
account core polarizations. If we take
the effect of core polarizations given by \cite{minamisono}, then
the resulting value comes out to be
 $\epsilon_{MEC}\approx 1.53\sim 1.59$ which should be compared
with the experimental value $\epsilon_{MEC}=1.57$. Given the drastic
simplification of the finite size effect of the system, this agreement should
not be taken too seriously. Nonetheless this indicates the correctness of
our theory.  One could certainly
do better by using a local density approximation in accounting for the
BR scaling in these light nuclei rather than using the average density.}
\be
\epsilon_{MEC}^{th}\approx 2.1\ \ \ (2.0)
\ee
in good agreement with the ``experimental" results of
\cite{warburton}, \cite{otherexp} and
\cite{minamisono}. Here the value in parenthesis is obtained with
the Landau formula (\ref{LACp}). The difference between the two
formulas (i.e., current algebra vs. Landau) is indeed small. This
is a check of the scaling of $f_\pi$ in combination with the
scaling of the Gamow-Teller constant $g_A$ in medium, discussed
below.
\subsection{The Gamow-Teller constant in nuclei}
\indent\indent
Recall that because of the pions which provide (perturbative)
non-local interactions to the Landau interaction, the Landau mass
for the nucleon scales differently from that of the vector mesons.
(See (\ref{BRscaling}) and (\ref{mstar2})). This difference is
manifested in the skyrmion description by the fact that the
coefficient of the Skyrme quartic term must also scale. In the
original discussion of the scaling based on the quark condensates
using the trace anomaly~\cite{BR91}, the Skyrme quartic term was
scale-invariant and hence the corresponding $g_A^\star$ was
non-scaling. So the scaling implied by (\ref{mstar2}) indicates
that the scaling of $g_A^\star$ is associated with the pionic
degrees of freedom.
This is consistent with the description based on the
Landau-Migdal $g_0^\prime$ interaction between a nucleon and a
$\Delta$ resonance~\cite{rho,ohtawakamatsu} and also with the QCD
sum-rule description of Drukarev and Levin~\cite{drukarev} who
attribute about 80\% of the quenching to the $\Delta-N$ effect.

If we equate the skyrmion relation~\cite{FR96,mr85}
\be
\frac{m_N^\star}{m_N}=\sqrt{\frac{g_A^\star}{g_A}}\Phi
\ee
to (\ref{mstar2}), we get
\be
\frac{g_A^\star}{g_A}=\left(1+\frac 13 F_1 (\pi)\right)^2
=\left(1-\frac 13 \tilde{F}_1 (\pi)\Phi\right)^{-2}.\label{quenching}
\ee
At nuclear matter density, this predicts\footnote{Since one expects
that when chiral symmetry is restored, $g_A$ will approach 1, it
may be thought that the evidence for $g_A^\star\approx 1$ in nuclei
is directly connected 
with chiral restoration. As
one of the authors (MR) has argued since a long time and as
mentioned above, this is not really the case. Neither in the
skyrmion picture nor in QCD sum rules is the quenching of $g_A$
simply related to a precursor behavior of chiral restoration. This
does not however mean that the quenching of $g_A$ carries no
information on the chiral symmetry restoration. As suggested
recently by Chanfray, Delorme and Ericson \cite{CDE}, if one were
to compute {\it all} pion-exchange-current graphs at one-loop order
that contribute to the in-medium $g_A$, the effect of
medium-induced change in the quark condesate would be largely
accounted for. In a way, this argument is akin to that for the
Cheshire-Cat (or dual) phenomenon we are advocating in the
description of the quark condensate in terms of quasiparticles.
Another issue that has generated lots of debate in the past and yet
remains confusing is the interpretation of an {\it effective}
constant $g_A^{eff}\approx 1$ actually observed in medium and heavy
nuclei. The debate has been whether the observed ``quenching" is
due to ``core polarizations" or ``$\Delta$-hole effect" (or other
non-standard mechanisms). Our view is that in the presence of BR
scaling, both are involved. In light nuclei in which the
Gamow-Teller transition takes place in low density, the tensor
force is mainly operative and the core polarization (i.e.,
multiparticle-multihole configurations) mediated by this tensor
force is expected to do  most of the quenching, while the
$\Delta$-hole effect directly proportional to density is largely
suppressed. In heavy nuclei, on the other hand, the tensor force is
quenched due to BR scaling, rendering the core polarization
mechanism ineffective while the increased density makes the
$\Delta$-hole effect dominant. What is seen in nature, in our view,
is the interplay between these two. It seems that this
complementarity is overlooked in the literature, due mainly to the
fact that the equivalence of various different approaches or the
Cheshire-Cat phenomenon is not yet fully appreciated. ``New
physics" will set in at truly large density, not in light nuclear
systems.}
\be
g_A^\star (\rho_0)\approx 1
\ee
and
\be
\frac{g_A^\star}{g_A}\approx \frac{f_\pi^\star}{f_\pi}=\Phi.
\ee
This is the relation we used in deriving (\ref{CAAC}). It should be
emphasized that this relation, being unrelated to the vacuum
property, cannot hold beyond $\rho\approx \rho_0$. Indeed as
suggested by the scaling given in \cite{BR91}, $g_A^\star
(\rho)\approx \alpha g_A$ with $\alpha$ a constant independent of
density, for $\rho\gsim \rho_0$. It would be a good approximation
to set $g_A^\star$ equal to 1 for $\rho\gsim\rho_0$.

The second form of (\ref{quenching}) shows that the quenching of
$g_A$ in matter is quite complex, both the pionic effect and the
vacuum condensate effect being confounded together. Again for the
reason given above, this relation cannot be extended beyond the
regime with $\rho\approx\rho_0$. We have no understanding of how
this formula and the $\Delta$-hole mechanism of
\cite{rho,ohtawakamatsu} are related. Our effort thus far has met
with no success. Understanding the connection would presumably
require the short-distance physics implied by both the
Landau-Migdal $g_0^\prime$ interaction and the Skyrme quartic term
(which is known to be more than just what results when the $\rho$
meson is integrated out of the chiral Lagrangian).
\subsection{Dropping meson masses}
\indent\indent
Our mapping of the scaling parameter $\Phi$ to a Landau
quasiparticle interaction at densities around nuclear matter
density gives evidence on the property of nucleons in medium.
However it should be recalled that we extracted the scaling
parameter $\Phi$ from the in-medium property of the vector mesons.
So the next question is: What evidence is there for the predicted
scaling in the meson masses ? There are some preliminary
experimental indications for the decrease in matter of the $\rho$
meson mass in recent nuclear experiments~\cite{evidence1,evidence2}
but we expect more definitive results from future experiments
at Gesellschaft f\"ur Schwerionenforschung (GSI) and other laboratories. 
In fact, this is
currently a hot issue in connection with the recent dilepton data
coming from relativistic heavy-ion experiments at Centre Europeen 
pour la Recherche Nucleaire (CERN).

When heavy mesons such as the vector mesons $\rho$, $\omega$ and
the scalar $\sigma$ are reinstated in the chiral Lagrangian, then
the mass parameters of those particles in the Lagrangian, when
written in a chirally invariant way, are supposed to appear with
asterisk and are assumed to scale according to
Eq.(\ref{BRscaling}). The question is: What is the physical role of
these mass parameters? If we assume that the mesons behave also
like quasiparticles, that is, like  weakly interacting particles
with the ``dropping masses," then physical observables will be
principally dictated by the tree diagrams of those particles
endowed with the scaling masses. In this case, the masses figuring
in the Lagrangian could be identified in some sense as ``effective"
masses of the particles in the matter. This line of reasoning was
used in the work of Li, Ko and Brown~\cite{LKB} to
interpret
the low-mass enhancement
of the Cherenkov Ring Electron Spectrometer (CERES) data~\cite{CERES}. 
As discussed in \cite{sbmr}, this
treatment is consistent with an effective Lagrangian which in the
mean field approximation yields
the nuclear
matter ground state as well as fluctuations around the ground
state. The parameters of the theory, as well as their density
dependence is determined by the properties of the ground state. The
work of \cite{sbmr,smr} shows that this scheme is
internally
consistent. However we
emphasize that the scaling we have established is for the mesons
that are highly off-shell and it may not be applied to mesons that
are near on-shell without further corrections (e.g., widths etc.).

The equivalence discussed above between the physics of the
``vacuum" property $\Phi$ and that of the  ``quasiparticle
interaction" $F_1$ due to the massive vector-meson degree of
freedom suggests that the ``bottom-up" approach -- going up in
density with a Lagrangian whose parameters are fixed at zero
density -- and the ``top-down" approach -- extrapolating with a
Lagrangian whose parameters are fixed at some high density -- can
be made equivalent at some intermediate point. If this is so in the
hot and dense regime probed by relativistic heavy-ion collisions,
then the CERES data should also be understandable in terms of
hadronic interactions without making reference to QCD variables.
Because of the complexity of the hadronic
descriptions, it will be difficult to relate the two directly but
the recent ``alternative explanation" of the CERES data in terms of
``melting of the vector mesons" inside nuclear matter manifested in
the increased width of the mesons due to hadronic
interactions~\cite{wambach} may be an indication for
the
``duality" we emphasized above. A possible
mechanism that could make the link between the two descriptions was
suggested recently by Brown et al.\ \cite{BLRRW}.
\section{Discussions}\label{discussions}
\indent\indent
What we have achieved in this paper can be summarized as follows.
\begin{enumerate}
\item
By means of nuclear response to electromagnetic convection current,
we have identified the BR-scaling parameter $\Phi$ with the scaling
nucleon mass $M^\star$. The Landau effective mass of the nucleon
$m_N^\star$ is in turn given in terms of $\Phi$ and the Goldstone
boson cloud of the broken chiral symmetry, i.e., pion, through the
parameter $\tilde{F}_1^\pi$.
The relation between
the orbital gyromagnetic ratio $\delta g_l$ and $m_N^\star$
provides the crucial link between $\Phi$ and the Landau parameter
$F_1^\omega$ coming from the massive degrees of freedom in the
isoscalar vector channel dominated by the $\omega$ meson.
\item The axial charge transitions in heavy nuclei provides a relation
between $\Phi$ and the in-medium pion decay constant
$f_\pi^\star/f_\pi$.
\item A Walecka-type linear model for nuclear matter with the
parameters of the Lagrangian scaling a'la Brown-Rho consistent with
chiral symmetry provides the connection between $\Phi$ and the
scaling of the vector-meson degrees of freedom ($\omega$ and
$\rho$) and scalar-meson degrees of freedom ``$\sigma$" in the
situation where the mesons are highly off-shell. This relation has
been 
checked against fluctuations into the
strangeness flavor, namely, kaon-nuclear interactions. So far the
check is only semi-quantitative and approximate but there is
consistency.
\item The relations verified up to nuclear matter density as described
above are extrapolated to high densities as in heavy-ion collisions
and neutron-star formation. How the scaling parameters extrapolate
beyond normal nuclear matter density is not predicted by theory and
should be deduced from lattice measurements and heavy-ion
experiments that are to come. Corrections to BR scaling as massive
mesons approach on-shell need be taken into account. The fit to the
available CERES data indicates however that the extrapolation to
higher density -- perhaps up to the chiral phase transition -- is
at least approximately correct under the conditions that prevail in
nucleus-nucleus collisions at SPS energies. How this could come
about was discussed in \cite{BLRRW}.
\item A rigorous derivation of BR scaling starting from an effective
chiral action via multiple scale decimations required for the
problem is yet to be formulated but the main ingredients, both
theoretical and phenomenological, seem to be available.
\end{enumerate}

So far, we have succeeded in mapping the chiral Lagrangian theory
with BR scaling to nonrelativistic Landau Fermi liquid theory. This
is natural since we worked in heavy-Fermion formalism for the
chiral Lagrangian field theory. However in order to apply the
correspondence to dense matter encountered in relativistic
heavy-ion collisions and in neutron stars, we should formulate the
mapping relativistically as in \cite{smr} where thermodynamic
properties of a BR-scaled chiral Lagrangian in the mean field were
shown to be consistent with the relativistic Landau formulas
derived by Baym and Chin~\cite{baymchin}. This work is in progress.

In discussing properties of dense matter, such as BR scaling of
masses and coupling constants, e.g., $f_\pi^\star$ , we have been
using a Lagrangian which preserves Lorentz invariance. This seems
to be at odds with the fact that the medium breaks Lorentz
symmetry. One would expect for instance that the space and time
components of a current would be characterized by different
constants. Specifically such quantities as $g_A$, $f_\pi$ etc.
would be different if they were associated with the space component
or time component of the axial current. So a possible question is:
How is the medium-induced symmetry breaking accommodated in the
formalism we have been discussing in this paper?

The answer to this question was provided in \cite{smr}. There the
argument was given in an exact parallel to Walecka mean field
theory of nuclear matter. One writes an effective Lagrangian with
all symmetries of QCD which in the mean field defines the
parameters relevant to the state of matter with density. The
parameters that become constants (masses, coupling constants etc.)
at given density are actually functionals of chiral invariant
bilinears in the 
nucleon fields. When the scalar
field $\phi$ and the bilinear ${\psi}^\dagger\psi$, where $\psi$ is
the nucleon field, develop a non-vanishing expextation value
Lorentz invariance is broken and the time and space components of a
nuclear current pick up different constants. This is how for
instance the Gamow-Teller constant $g_A$ measured in the space
component of the axial current is {\it quenched} in medium while
the axial charge measured in the axial charge transitions is {\it
enhanced} as described in Section \ref{test}. If one were to
calculate the pion decay constant in medium, one would also find
that the quantity measured in the space component is different from
the time component. The way Lorentz-invariant Lagrangians figure in
nuclear physics is is in some sense similar to what happens in
condensed matter physics. For example, on a lattice where there is
not even rotational invariance, one finds a Lorentz-invariant
dispersion formula. Another well-known example is the fractional
quantized Hall effect which is described by a Lorentz-invariant
Lagrangian containing the Chern-Simons term \cite{zee}.

\subsection*{Acknowledgments}
\indent\indent
One of us (CS) is grateful to Dong-Pil Min for useful comments.
MR acknowledges the hospitality of Theory Group of GSI
where part of this work was done under the support of
the Humboldt Foundation
and that of Korea Institute for Advance Study (KIAS)
where part of this paper was written.
The work of CS was supported
in part by
KOSEF through the Center for Theoretical Physics, Seoul National University,
and in part by Korea Ministry of Education (Grant No. BSRI98-2418).

\pagebreak

\setcounter{equation}{0}
\setcounter{figure}{0}
\renewcommand{\theequation}{\mbox{A.\arabic{equation}}}
\renewcommand{\thefigure}{\mbox{A.\arabic{figure}}}
\section*{Appendix A: Relativistic Calculation of $F^\pi_1$}
\begin{figure}[htb]
\centerline{\epsfig{file=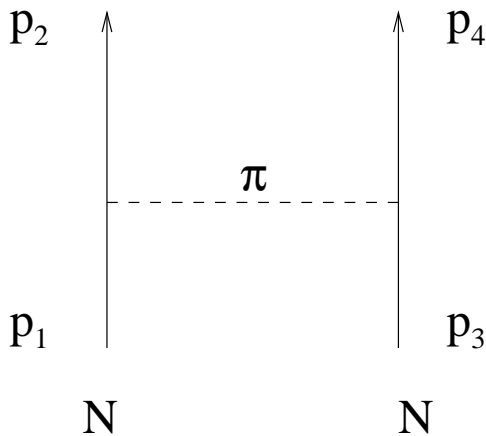,width=2.5in}}
\caption[f3]{\small The-one-pion-exchange diagram that gives rise to
$F_1^\pi$. }\label{fpi}
\end{figure}

\indent\indent
In the text, the Landau parameter $F_1^\pi$ (or $f^\pi$) was calculated
nonrelativistically via the Fock term of Figure \ref{fpi}.
Here we calculate it relativistically
by Fierz-transforming the one-pion-exchange graph and
taking the Hartree term. This procedure is important for implementing
relativity in the connection between Fermi-liquid theory and chiral Lagrangian
theory along the line discussed by Baym and Chin~\cite{baymchin}.

The one-pion-exchange potential in Figure \ref{fpi} is
\be
V_{\pi}= -g_{\pi NN}^2(\taub_{21}\cdot\taub_{43})
\frac{\bar{u}_2\gamma^5u_1\bar{u}_4\gamma^5
u_3}{(p_2-p_1)^2-m_\pi^2}.
\ee
The Dirac spinors are normalized by
\be
u^\dagger (p,s)u(p,s^\prime )=\delta_{ss^\prime}.
\ee
By a Fierz transformation, we have
\be
\taub_{21}\cdot\taub_{43}
=\frac12 (3\delta_{41}\delta_{23}-\taub_{41}\cdot\taub_{32})
\label{fierzt}\ee
and
\be
\bar{u}_2\gamma^5u_1\bar{u}_4\gamma^5u_3
&=&\frac{1}{4}[\bar{u}_4u_1\bar{u}_2u_3-
\bar{u}_4\gamma^\mu u_1\bar{u}_2\gamma_\mu u_3\label{fierzf}\\ & &+
\bar{u}_4\sigma^{\mu\nu}u_1\bar{u}_2\sigma_{\mu\nu}u_3+
\bar{u}_4\gamma^\mu\gamma^5u_1\bar{u}_2\gamma_\mu\gamma^5u_3+
\bar{u}_4\gamma^5u_1\bar{u}_2\gamma^5u_3].\nonumber
\ee
Remembering a minus sign for the fermion exchange, we obtain the
corresponding pionic contribution to the quasiparticle interaction
at the Fermi surface, $f^\pi = -V_\pi (\pb_1=\pb_4=\pb,
\pb_2=\pb_3=\pb^\prime ,\pb^2 =\pb^{\prime 2}=k_F^2)$ (see (\ref{qpint})).
Decomposing $f^\pi$ as
\be
f^\pi =\frac{3-\taub\cdot\taub^\prime }{2}
(f_S+f_V+f_T+f_A+f_P)\label{fss}
\ee
where $S$, $V$, $T$, $A$ and $P$ represent scalar, vector, tensor,
axial vector and pseudoscalar channel respectively, we find
\be
f_S&=&-\frac{m_N^4f^2}{E_F^2m_\pi^2}\frac{1}{q^2+m_\pi^2}
\nonumber\\
f_V&=&\frac{m_N^4f^2}{E_F^2m_\pi^2}\frac{1}{q^2+m_\pi^2}
\left( 1+\frac{q^2}{2m_N^2} \right) \nonumber\\
f_T&=&-\frac{m_N^4f^2}{E_F^2m_\pi^2}\frac{1}{q^2+m_\pi^2}
\left( \sigmab\cdot\sigmab^\prime (1+\frac{q^2}{2m_N^2})
+\frac{2\sigmab^\prime\cdot\pb\sigmab\cdot\pb^\prime
-\sigmab\cdot\pb\sigmab^\prime\cdot\pb
-\sigmab\cdot\pb^\prime\sigmab^\prime\cdot\pb^\prime}{2m_N^2}
\right)\nonumber\\
f_A&=&\frac{m_N^4f^2}{E_F^2m_\pi^2}\frac{1}{q^2+m_\pi^2}
\left( \sigmab\cdot\sigmab^\prime
-\frac{2\sigmab\cdot\pb\sigmab^\prime\cdot\pb^\prime
-\sigmab\cdot\pb\sigmab^\prime\cdot\pb
-\sigmab\cdot\pb^\prime\sigmab^\prime\cdot\pb^\prime}{2m_N^2}
\right)\nonumber\\
f_P&=&0.\label{decomp}
\ee
with $E_F =\sqrt{k_F^2+m_N^2}$ and $q=|\pb -\pb^\prime |$.
Thus we obtain
\be
f^\pi
&=&\frac{f^2}{m_\pi^2}\frac{m_N^2}{E_F^2}\frac{1}{q^2+m_\pi^2}
\left( \sigmab\cdot\qb\sigmab^\prime\cdot\qb
-\frac{q^2(1-\sigmab\cdot\sigmab^\prime )}{2}\right)
\frac{3-\taub\cdot\taub^\prime }{2}\label{rfpi}\\
&=&\frac13\frac{f^2}{m_\pi^2}\frac{m_N^2}{E_F^2}\frac{q^2}{q^2+m_\pi^2}
\left( 3\frac{\sigmab\cdot\qb\sigmab^\prime\cdot\qb}{q^2}
-\sigmab\cdot\sigmab^\prime
+\frac12 (3-\sigmab\cdot\sigmab^\prime)\right)
\frac{3-\taub\cdot\taub^\prime }{2}.\nonumber
\ee
In the nonrelativistic limit, $E_F
\sim m_N$ and we recover (\ref{vpi}). The factor $m_N/E_F$ comes since
there is one particle in the unit volume which decreases
relativistically as the speed increases. Note that only $f_S$ and
$f_V$ in (\ref{decomp}) are spin-independent and contribute to
$F_1^\pi$. The $f_S$ is completely canceled by the leading term of
$f_V$ with the remainder giving $F_1^\pi$. In this way of deriving
the Landau parameter $F_1$, it is the vector channel that plays the
essential role.

\setcounter{equation}{0}
\setcounter{figure}{0}
\renewcommand{\theequation}{\mbox{B.\arabic{equation}}}
\section*{Appendix B: Particle-Hole Contribution to the Vector Current}
\indent\indent
The leading contribution of the particle-hole polarization with one-pion
exchange is shown in Figure \ref{ph}. This graph was computed by several
authors (e.g., see \cite{bentz}) and is given in the limit
$\omega /q\rightarrow 0$ by
\be
\Jvec_{ph}=-\sum_{\tau^\prime}
\la \taub (1)\cdot\frac{1+\tau_3^\prime}{2}\taub (2)\ra
\int\frac{d^3p}{(2\pi )^3}\hat{\pb}\delta (k_F-|\pb |)f^\pi_s
\ee
where $f_s^\pi \equiv f_S+f_V+f_T+f_A+f_P$.
The isospin factor is given by the Fierz transformation:
\be
\sum_{\tau^\prime}
\la\taub (1)\cdot\frac{1+\tau_3^\prime}{2}\taub (2)\ra
&=&\sum_{\tau^\prime}
\la\frac34 -\frac14\taub\cdot\taub^\prime
+\frac34 \tr [\tau_3^\prime ]-\frac14\taub\cdot\tr [\tau_3^\prime
\taub^\prime ]\ra\nonumber\\
&=&\frac32-\frac12\tau_3.\label{isoint}
\ee
Note that the factor $\frac32$ comes from $f_\pi$ and $\frac12\tau_3$ from
$f^\prime_\pi$. In the limit that we are concerned with
(i.e., $T=0$ and $\omega /q\rightarrow 0$),
we find
\be
\Jvec_{ph}
&=&-\frac{1}{3\pi^2}\kbhat k_F^2(f_1+f_1^\prime \tau_3)\\
&=&-\frac{\Bk}{m_N}
\frac{\tilde{F}_1 (\pi)+\tilde{F}_1^\prime (\pi)\tau_3}{6}.\nonumber
\ee
Contributions from heavy-meson exchanges are calculated in a similar way.
\pagebreak

\end{document}